\documentclass[pdftex, copyright, creativecommons, noderivs, noncommercial]{eptcs}
\providecommand{\event}{ESSS 2014}

\newif\ifUseTikzExternalPictures
\UseTikzExternalPicturestrue

\usepackage[utf8]{inputenc}

\usepackage{alloy}
\usepackage{subfigure}
\usepackage{wrapfig}
\usepackage{relsize}

\usepackage{graphicx}
  \graphicspath{{./pictures/}{./pictures/tikzExternal/}}
  \DeclareGraphicsExtensions{.pdf,.png}

\ifUseTikzExternalPictures
\else
  \usepackage{tikz}
    \usetikzlibrary{fit}
    \usetikzlibrary{shapes.symbols} 
    \usetikzlibrary{intersections}
    \usetikzlibrary{backgrounds}
    \usetikzlibrary{through}
    \usetikzlibrary{patterns}
    \usetikzlibrary{positioning}
  
  \usetikzlibrary{external}
  \tikzsetexternalprefix{pictures/tikzExternal/}
  \tikzsetfigurename{\tikzexternalrealjob_F}
  \tikzset{external/force remake}
\fi

\title{
  In my Wish List, an Automated Tool for Fail-Secure \\ Design Analysis: an Alloy-Based Feasibility Draft
}
\author{
  Gurvan Le Guernic
  \institute{
    DGA Maîtrise de l'Information \\
    35998 Rennes Cedex 9, France
  }
  \email{Gurvan.Le-Guernic@intradef.gouv.fr}
}
\def\titlerunning{Fail-Secure Design Analysis: an Alloy-Based Feasibility Draft}
\def\authorrunning{G. Le Guernic}

\newcommand{\NAMILCOM}[0]{\textsc{NAMilCom}}

\begin{document}

\maketitle

\begin{abstract}
 A system is said to be fail-secure, sometimes confused with fail-safe, if it maintains its security requirements even in the event of some faults.
 Fail-secure analyses are required by some validation schemes, such as some Common Criteria or NATO certifications. However, it is an aspect of security which as been overlooked by the community. This paper attempts to shed some light on the fail-secure field of study by: giving a definition of fail-secure as used in those certification schemes, and emphasizing the differences with fail-safe; and exhibiting a first feasibility draft of a fail-secure design analysis tool based on the Alloy model checker.
\end{abstract}


\section{Introduction}

 Security is an information system property of continuously growing importance.
 It is widely studied by academics, taken into account by industries in the design and implementation of their systems and equipments, and a concern for customers. However, it is a property which encompasses some subproperties which seem to attract less attention from the academic community and general purpose industries, but are of high interest for some specific industries. \emph{Fail-secure} behavior is such a subproperty.

 Fail-secure, which is sometimes confused with the concept of fail-safe, is the property of a system or equipment to maintain its \emph{security} properties even in the event of \emph{some} faults. Section~\ref{sec:fail-secure-analysis} defines this concept for the purpose of this paper and some evaluation schemes, such as the Common Criteria (CC)~\cite{CC-3.1} or North Atlantic Treaty Organization (NATO) Military Committee (\NAMILCOM{}) certifications~\cite{NC3B:2009:tidcscm}. 
 In addition to be required for some of those evaluations, fail-secure design analyses are welcome in the early stages of the engineering of fail-secure systems. Indeed, making a system fail-secure imposes additional constraints on the functional and physical architectures of the system. At the certification stage, it is often difficult to modify a product in order to take into account those constraints, particularly concerning its physical architecture, which advocates for the need of early fail-secure design analyses.

 This aspect of security, fail-secure behavior, has not attracted much interest from the academic community. Therefore, there are very few tools and techniques for fail-secure analysis of software or hardware architectures (Sect.~\ref{subsec:related-work}).
 Nowadays, real world fail-secure software or hardware architectures analyses 
 are mostly carried out manually. Hence, such analyses are time consuming and their quality depends heavily on the experience of the analyst. Section~\ref{sec:alloy-based-feasibility-draft} presents a first attempt at using the Alloy ``model finder'' to build a tool for automatic fail-secure analysis of early designs. This first trial is not a prototype yet, but rather a draft demonstrating the apparent feasibility of the approach. As stated in the conclusion in Sect.~\ref{sec:conclusion}, there is much more work to accomplish in this niche, which seems to offer the opportunity for some challenging academic work.

\section{Fail-Secure Analysis} \label{sec:fail-secure-analysis}

 In some communities, there is a lot of confusion between the terms ``fault-tolerant'', ``fail-safe'' and ``fail-secure''. This confusion is detrimental to attracting attention to the fail-secure design and analysis problems, and impedes understanding and communication.
 This section tries to clarify the meaning of the term ``fail-secure'' as used in this paper and
 other contexts\footnote{\url{http://en.wikipedia.org/wiki/} [ \href{http://en.wikipedia.org/wiki/Electromagnetic_lock}{\nolinkurl{Electromagnetic_lock}}$\ \mid\ $\href{http://en.wikipedia.org/wiki/Fail-safe}{\nolinkurl{Fail-safe}}$\ \mid\ $\href{http://en.wikipedia.org/wiki/Life-critical_system}{\nolinkurl{Life-critical_system}} ]}, such as the Common Criteria~\cite{CC-3.1}.

 \begin{wrapfigure}{r}{.45\linewidth}
   \vspace*{-2ex}

   \ifUseTikzExternalPictures\else%
     \input{tikz_common_safety-security}
   \fi%
   \hspace*{\stretch{1}}
   %
     \subfigure[Safety]{%
       \ifUseTikzExternalPictures%
         \includegraphics{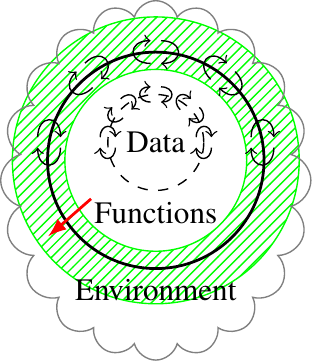}
       \else%
         \input{tikz_safety}%
       \fi%
     }
   %
   \hspace*{\stretch{1}}
   %
     \subfigure[Security]{%
       \ifUseTikzExternalPictures%
         \includegraphics{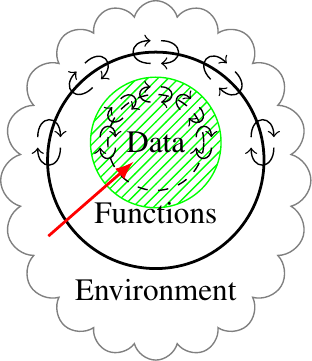}
       \else%
         \input{tikz_security}%
       \fi%
     }
   %
   \hspace*{\stretch{1}}

   \hspace*{\stretch{1}}
   \ifUseTikzExternalPictures%
     \includegraphics{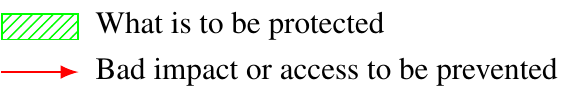}
   \else%
     \input{tikz_common_safety-security_legend}
   \fi%
   \hspace*{\stretch{1}}

   \vspace*{-1ex}

   \caption{Safety versus security}
   \label{fig:safety-vs-security}

   \vspace*{-0.5ex}
 \end{wrapfigure}

\begin{wrapfigure}{L}{.515\linewidth}
   \ifUseTikzExternalPictures%
     \vspace*{-6ex}
   \else%
     \vspace*{-6ex}
   \fi%

   \centering
       \ifUseTikzExternalPictures%
         \includegraphics{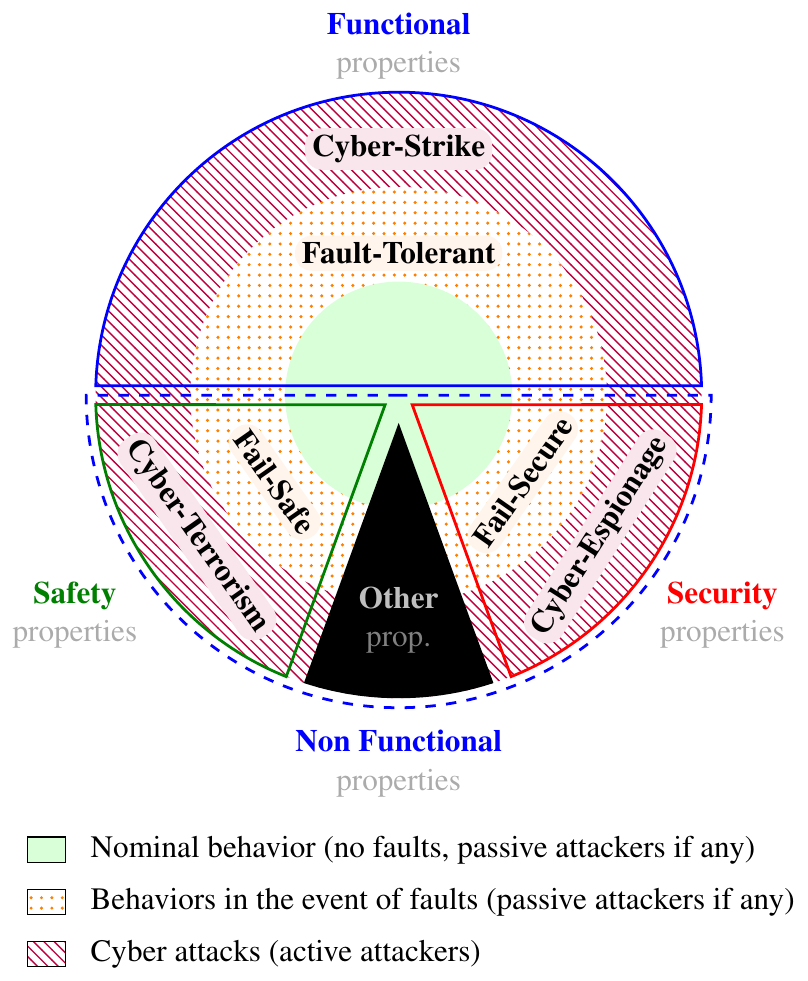}
       \else%
         \input{tikz_nomenclature}
       \fi%

   \ifUseTikzExternalPictures%
     \vspace*{-1ex}
   \else%
     \vspace*{-1ex}
   \fi%

   \caption{Fault-tolerant, fail-safe and fail-secure}
   \label{fig:nomenclature}

   \vspace*{-8ex}
 \end{wrapfigure}

 Figure~\ref{fig:safety-vs-security} depicts, in those contexts, the principal meanings of the terms ``safety'' and ``security'', on which are built ``fail-safe'' and ``fail-secure'' meanings. It pictures a system whose
 behavior (functions)
 influences and is influenced by its
 data on one side, and interacts with the environment on the other. ``Safety'' is mainly concerned by the protection of the environment and some of the system's functionalities. Safety properties ensure that the system's behavior does no ``harm'' to its environment or the system itself. ``Security'' relates mainly to the protection of the (sensitive) data, ensuring that no ``villainy'' from the environment impacts those data.
 Adding faults, a system is ``fault-tolerant'' if its \emph{functional requirements} hold even in the event of faults; and then degrade proportionally to the number of faults (there is no single point of system-wide failure). A system is ``fail-safe'' if its \emph{safety requirements} hold in the event of faults, and ``fail-secure'' if its \emph{security requirements} hold in the event of faults.
 Additionally, all those terms relate to the behavior of the system in the event of faults (and presence of passive attackers if any), but not to its behavior under active attacks. Resistance to active attackers is yet another concept with its own fields of studies. Those distinctions are summarized in Fig.~\ref{fig:nomenclature}.

\paragraph{Analogy with electromagnetic lock.} An electromagnetic lock is said to be fail-safe with regard to power failure if it is in the unlocked position when no power is applied to it. While it is fail-secure if it is in the locked position when no power is applied. With the appropriate organizational procedures, assuming that the safety requirements are that employees can always escape and security requirements are that secret documents in safes can not be taken outside, a building can be fail-safe \emph{and} fail-secure with regard to power failure if its office doors are fail-safe and safe doors are fail-secure. In case of power outage, the employees can escape the building and the secret documents are locked inside the safes.

\paragraph{Fail-secure according to the Common Criteria.} The Common Criteria are an international standard (ISO/IEC 15408)~\cite{CC-3.1}. This standard is a framework for the security certification of computer information systems. It allows vendors to select functional and assurance requirements that their products meet. Those claims are then verified by independent organizations.
 One of those functional requirements, FPT\_FLS.1~\cite[Part 2 (p.~128)]{CC-3.1} \texttt{Failure with preservation of secure state}, deals with the fail-secure property. It states ``that the TSF
 [(Target Of Evaluation Security Functionality)]
 preserve a secure state in the face of the identified failures''. The vendor shall state with regard to which failures (power loss, memory error, \dots) his product is fully fail-secure; i.e. no security requirement is broken in case of any of those failures.
%
%
 There is no obligation to include FPT\_FLS.1. However it appears in many protection profiles (PP) 
 and security targets (ST)
, such as the 2011 Trusted Computing Group's PP for TPM~\cite{TCG:2011:PP_TPM} or the 2000 ST of the Naval Research Laboratory's Network Pump~\cite{moore:2000:ST-NP}.

\paragraph{Fail-secure according to the \NAMILCOM{} certification.} In order to enter the NATO market and deal with sensitive NATO data, security equipments need to get a \NAMILCOM{} certification. Part of this certification process~\cite{NC3B:2009:tidcscm} requires a fail-secure analysis of the equipment. The goal of this analysis~\cite{DGA-MI:2012:guide-SeD_v4} is not to verify against which failures the equipment is fully fail-secure (not even one security requirement violated), as for the Common Criteria. Rather, it is to verify that the minimum number of faults before sensitive data is potentially compromised (leaked to the outside) is high enough. This number depends on the classification level of the data leaked and involves a notion of functional and physical independence between faults that is not addressed in this paper.

\subsection{Related work} \label{subsec:related-work}

 To the best of our knowledge, there is really little work on fail-secure design analysis.
 Rae and Fidge~\cite{rae:2005:ifa4fsd} propose an analysis based on matrices composition. The information flows inside a given atomic component are described in a single matrix (either for one or all operating modes, including different fault behaviors). Those matrices are then ``multiplied'' following the design of the system to get the overall information flows. This analysis follows the CC approach, specifying the faults taken into account, rather than the \NAMILCOM{} approach which is impervious to the precise type of faults. On the downside, this approach does not scale well.
 To the exception of this work, there seems to be no work addressing directly the fail-secure design analysis problem.
 Widening the scope of related works, an information flow analysis taking into account failures could be adapted to provide a fail-secure design analysis for some types of confidentiality and integrity requirements, which are the most common security requirements --- excluding access control. However, even if there are lots of information flow analyses~\cite{sabelfeld+03lbifs,leguernic07thesis}, to the best of our knowledge, the only information flow analyses handling component failures are those of the \textsc{Sifa} tool~\cite{mccomb:2005:SIFA} and its extension to micro-controller software~\cite{mills:2012:tsdasced}. \textsc{Sifa} embeds the notion of \emph{faults} into a more generic notion of \emph{mode}. However, \textsc{Sifa} is a circuitry-oriented information flow analysis mainly aimed at evaluating final products. It may therefore not be best suited for a fail-secure analysis of functional and physical architecture designs at the (detailed) specification level.

\section{An Alloy-based Feasibility Draft} \label{sec:alloy-based-feasibility-draft}

\lstnewenvironment{myAlloy}[1][]{
  \lstset{
    language=alloy,
    floatplacement={tbp},captionpos=b,
    xleftmargin=3em,xrightmargin=2em,
    breaklines=true, morecomment=[is]{/**}{**/},
    basicstyle=\smaller,
    mathescape,
    frame=single, frameround=tttt,
    numbers=left,
    firstnumber=last, 
    escapeinside={(*@}{@*)},
    #1}
}{}

\begin{figure}[!htb]
  \centering
%
    \includegraphics[width=\linewidth,clip=true,trim=1cm 1.1cm 1cm 1.5cm]{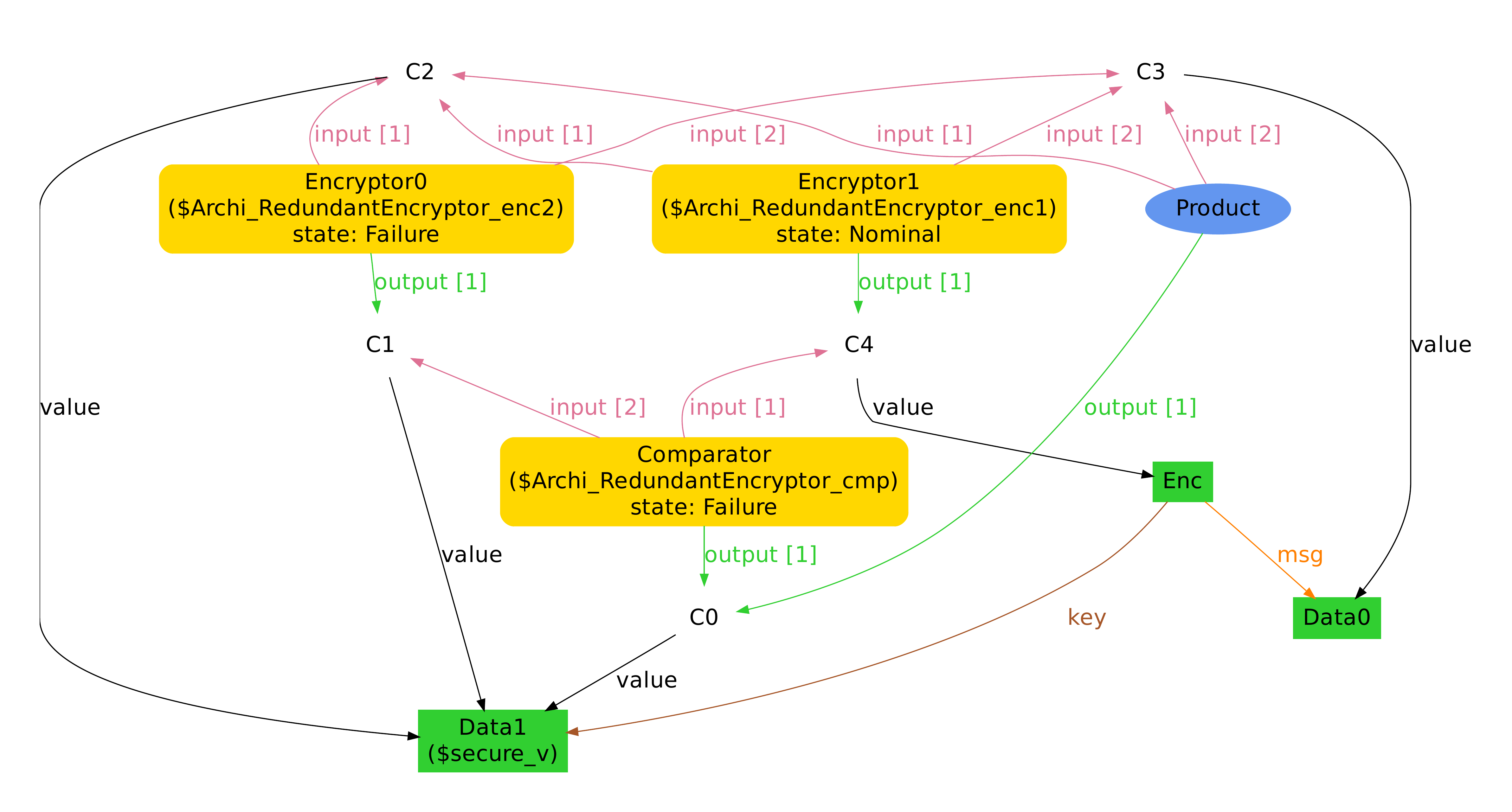}
%
  \caption{Result of the fail secure analysis against 2 faults for the redundant architecture}
  \label{fig:draft-result}
\end{figure}

 This section explores the possibility to build a fail-secure (according to \NAMILCOM{}) design analysis tool based on Alloy~\cite{jackson:2012:salla,Jackson:2002:alloy-alomn}. As an experiment, Alloy is used to verify if the high level architecture design of a fictitious encrypting product is fail-secure against 1 or 2 faults. This product takes as input a key and a message and returns as output the message encrypted by the key. Its security requirement (that is to be preserved even in presence of faults) states that ``the product should never directly output the key or message''. The redundant architecture of the product consists in 2 encrypting components followed by a comparator, as seen in Fig.~\ref{fig:draft-result}. The comparator outputs one of its inputs iff its two inputs are the same.

 Alloy is sometimes coined a ``model finder'' by its authors. It can also be seen as a ``meta-model checker''. The Alloy language is used to state constraints that describe a set of models, or structures. It is also used to define properties that are verified with the \emph{Alloy Analyzer}. For a given property (predicate or assertion), this tool finds models, in the provided set, for which this property holds; or states that there is no model, corresponding to the meta-model provided, for which the property holds.

 The approach followed by this feasibility draft is to provide a framework that defines a ``running'' product architecture as a set of components, defined as a function in a given state, connected by connectors (List.~\ref{lst:framework}, lines \ref{lst:sig_connector} to \ref{lst:sig_component}). Each connector is associated to a symbolic value. The state of a component is either \emph{nominal} or in \emph{failure} (
 lines \ref{lst:sig_functionalState} to \ref{lst:sig_nominal-failure}%
). Intrinsically, a component is a function that relates possible output values to input values. For this simplified draft, if a component is in failure state, then its only \emph{functional} constraint (
 line \ref{lst:cst_failureBehavior}) is that its output values belong to its input values (for each output, the value of one input is directly transferred).

\begin{myAlloy}[caption={Framework}, label={lst:framework}]
sig Connector {	value: one Value }{} (*@\label{lst:sig_connector}@*)
abstract sig Function { input: Int -> lone Connector, output: Int -> lone Connector, $\cdots$ }
  { $\cdots$ }
abstract sig Component extends Function { state: one FunctionalState } (*@\label{lst:sig_component}@*)
  { state = Failure implies Outputs.value in Inputs.value } (*@\label{lst:cst_failureBehavior}@*)
one sig Product extends Function {}{ $\cdots$ }
abstract sig FunctionalState {} (*@\label{lst:sig_functionalState}@*)
one sig Nominal, Failure extends FunctionalState {} (*@\label{lst:sig_nominal-failure}@*)
$\cdots$
\end{myAlloy}

 Instantiating this framework consists in defining components, such as the comparator in List.~\ref{lst:instantiation}, and describing an architecture based on those components. The remaining of this section is based on an instantiation of this framework for
 the redundant architecture of the fictitious encrypting product described above.
 This architecture corresponds to the yellow rounded boxes in Fig.~\ref{fig:draft-result}. A comparator has two inputs, \texttt{input[1]} and \texttt{input[2]}, and one output (List.~\ref{lst:instantiation}, line \ref{lst:sig_comparator-IO}). Its nominal behavior is to return \texttt{Null} if \texttt{input[1]} and \texttt{input[2]} differ, and return the value of \texttt{input[1]} otherwise (lines \ref{lst:sig_comparator-behavior-start}--\ref{lst:sig_comparator-behavior-end}). ``Encryptors'' are similarly defined to return the encryption of \texttt{input[2]} with the key in \texttt{input[1]}. The fictitious product is an encryptor with a redundant architecture (line \ref{lst:pred_RedundantEnc_start}): it has 2 inputs and 1 output (line \ref{lst:pred_RedundantEnc_IO}); it is composed of 2 encryptors and 1 comparator (line \ref{lst:pred_RedundantEnc_components}); it shares its inputs with the 2 encryptors \texttt{enc1} and \texttt{enc2} (lines \ref{lst:pred_RedundantEnc_shareI1}--\ref{lst:pred_RedundantEnc_shareI2}); the outputs of \texttt{enc1} and \texttt{enc2} are internally connected to the inputs of the comparator \texttt{cmp} (line \ref{lst:pred_RedundantEnc_internalConnections}); and it shares its output with \texttt{cmp}
 (line \ref{lst:pred_RedundantEnc_shareO}).

\begin{myAlloy}[caption={Architecture instantiation}, label={lst:instantiation}]
sig Comparator extends Component {}{ (input.Connector = 1 + 2) && (output.Connector = 1) (*@\label{lst:sig_comparator-IO}@*)
  state in Nominal implies ( equalValues[input[1].value, input[2].value] (*@\label{lst:sig_comparator-behavior-start}@*)
    implies (output[1].value = input[1].value) else output[1].value in Null ) (*@\label{lst:sig_comparator-behavior-end}@*)
}
$\cdots$
pred Archi_RedundantEnc { (*@\label{lst:pred_RedundantEnc_start}@*)
  Product.input.Connector = 1+2 && Product.output.Connector = 1 (*@\label{lst:pred_RedundantEnc_IO}@*)
  #Encryptor = 2 && one Comparator && Component = Encryptor + Comparator (*@\label{lst:pred_RedundantEnc_components}@*)
  some disj enc1,enc2:Encryptor, cmp:Comparator |
    Product.input[1] = enc1.input[1] && Product.input[1] = enc2.input[1] && (*@\label{lst:pred_RedundantEnc_shareI1}@*)
    Product.input[2] = enc1.input[2] && Product.input[2] = enc2.input[2] && (*@\label{lst:pred_RedundantEnc_shareI2}@*)
    cmp.input[1] = enc1.output[1] && cmp.input[2] = enc2.output[1] && (*@\label{lst:pred_RedundantEnc_internalConnections}@*)
    Product.output[1] = cmp.output[1] (*@\label{lst:pred_RedundantEnc_shareO}@*)
} (*@\label{lst:pred_RedundantEnc_end}@*)
\end{myAlloy}

 Verifying that this redundant architecture is fail-secure to a given number of component failures consists in checking a relatively simple assertion.
 The security requirement stated at the beginning of the section is encoded as ``no input value is directly output'' (List.~\ref{lst:analysis}, line \ref{lst:pred_secure})).
 A product is fail-secure up to \texttt{n} faults iff: it is not running securely only if
 more than \texttt{n} components are in a failure state (line \ref{lst:pred_fail-secure}).

\begin{myAlloy}[caption={Fail-secure analysis}, label={lst:analysis}]
pred secure { no v:Product.Outputs.value | v in Product.Inputs.value } (*@\label{lst:pred_secure}@*)
pred failsecureToN [n:Int] { (not secure) implies #(state.Failure) > n } (*@\label{lst:pred_fail-secure}@*)
assert redundantFailsecureToOne { (Archi_RedundantEnc and $\cdots$) implies failsecureToN[1] } (*@\label{lst:ass_fail-secure-to-1}@*)
assert redundantFailsecureToTwo { (Archi_RedundantEnc and $\cdots$) implies failsecureToN[2] } (*@\label{lst:ass_fail-secure-to-2}@*)
\end{myAlloy}

 The redundant architecture of the fictitious product (corresponding to the yellow rounded boxes in Fig.~\ref{fig:draft-result}) is fail-secure to 1 fault.
 If a single component is in failure state, whatever the fault, the overall product can not leak its inputs. If it is one of the two encryptors which is in failure, the comparator receives two different values as input and returns \texttt{Null}. If it is the comparator which is in failure, then the two encryptors are not faulty and provide the same encrypted values to the comparator which can only output one of its inputs. As expected, Alloy is unable to find a counter-example when asked to verify the assertion corresponding to the redundant architecture being fail-secure up to 1 fault (line \ref{lst:ass_fail-secure-to-1}).
 However, this redundant architecture is not fail-secure to 2 faults. Figure~\ref{fig:draft-result} shows a counter-example returned by Alloy when asked to verify the assertion corresponding to the redundant architecture being fail-secure up to 2 faults (line \ref{lst:ass_fail-secure-to-2}). In this counter-example, there is a fault in the comparator which, as a result, returns the erroneous value returned by the first encryptor (which is also in failure state and returns directly the encryption key).

\section{Conclusion} \label{sec:conclusion}

 This framework demonstrates the feasibility to encode into Alloy the main aspects of a fail-secure design analysis. Of course, there is much more work to do for this framework to be usable. First, this framework should integrate the notion of functional and physical independence of faults, and refine the definition of ``secure'' to reflect the dependence to the classification of data. In a longer run, the Alloy model should be automatically extracted from models expressed in more standard languages, such as UML or SysML.

 The development of an automated tool for fail-secure design analysis is a niche market. However, it would be highly beneficial to architects having to do a fail-secure analysis, which is a requirement for dealing with sensitive NATO data. Due to the challenges that need to be resolved and the lack of contenders, the field of fail-secure analysis is an interesting application area for academic work.

\bibliographystyle{eptcs}
\bibliography{2014-02_alloy-and-failsecure}

\end{document}